\documentclass[a4paper,11pt]{article}
\usepackage{pos}

\usepackage[switch]{lineno}
\def\nudoubt{NuDoubt${}^{++}$}

\usepackage{hyperref}
\stepcounter{footnote}
\usepackage{subcaption}
\usepackage{orcidlink}

\renewcommand{\logo}{\relax}

\title{Search for Double Beta Plus Decays with NuDoubt${}^{++}$}

\author*[a]{Stefan~Schoppmann\,\orcidlink{0000-0002-7208-0578}\,}
\onbehalf{on behalf of the NuDoubt${}^{++}$ Collaboration\footnote{Website: \url{https://nudoubt.uni-mainz.de}}}

\affiliation[a]{Johannes~Gutenberg-Universität~Mainz, Detektorlabor, Exzellenzcluster~PRISMA${}^+$, 55128~Mainz, Germany}

\emailAdd{nudoubt@lists.uni-mainz.de}
\emailAdd{stefan.schoppmann@uni-mainz.de}

\abstract{Double beta plus decay is a rare nuclear disintegration process.
Difficulties in its measurement arise from suppressed decay probabilities, experimentally challenging decay signatures and low natural abundances of suitable candidate nuclei.
In this presentation, we propose \nudoubt, a new detector concept to overcome these challenges.
It is based on the first-time combination of hybrid and opaque scintillation detector technology paired with novel light read-out techniques.
This approach is particularly suitable detecting positron (beta plus) signatures.
We expect to discover two-neutrino double beta plus decay modes within 1 tonne-week exposure and are able to probe neutrinoless double beta plus decays at several orders of magnitude improved significance compared to current experimental limits.}

\FullConference{Draft submitted to: 19th International Conference on Topics in Astroparticle and Underground Physics (TAUP2025)\\
24–30 Aug 2025\\
Xichang, China\\}

\begin{document}
\maketitle

\section{Introduction}
An open question in particle physics is regarding the nature of the neutrino.
While all known charged elementary fermions are Dirac particles, the neutrino could be the only Majorana fermion in the Standard Model (SM), i.e.~identical to its own anti-matter partner~\cite{Majorana_1937}.
The discovery of this Majorana property would revolutionise our understanding of our universe, because it yields a mechanism for lepton number violation (LNV).
LNV in turn can lead to an explanation for the matter/anti-matter asymmetry in the early universe by causing leptogenesis and baryogenesis, i.e.~the creation of leptons and baryons without their anti-matter partners that resulted in today's predominance of matter over anti-matter.

Neutrino-less double beta decay searches represent the most sensitive experimental way to probe the Majorana nature of the neutrino~\cite{Furry_1939}.
Double beta ($\beta\beta$) decay is a rare nuclear disintegration which changes the nuclear charge number $Z$ by two units while keeping the nucleon number $A$ unchanged~\cite{Goeppert-Mayer_1935}.
In regular double beta decays ($2\nu\beta\beta$), two neutrinos are ejected, while in neutrino-less decays ($0\nu\beta\beta$) the neutrinos are internally absorbed leading to LNV.
Currently, searches for $0\nu\beta\beta$ and $2\nu\beta\beta$ focus on negative double beta decays, i.e.~double electron emission, and have found evidence for $2\nu2\beta^-$, but none for $0\nu2\beta^-$.
In contrast, of the positive double weak decays, i.e.~double positron decay ($2\beta^+$), positron emitting electron capture (EC$\beta^+$), and double electron capture (2EC) neither $2\nu2\beta^+$ nor $2\nu\text{EC}\beta^+$ nor any neutrino-less decays have been seen, even though $2\nu2\beta^+$ and $2\nu\text{EC}\beta^+$ are undisputedly expected to exist in nature.
Their measurement is particularly difficult due to suppressed decay probabilities in terms of phase space, challenging background suppression, and low natural abundances of suitable candidate nuclei.

\section{Technology Concepts}
To perform the first every measurements of positron-emitting double weak decays, we are going to exploit the combination of four novel technology concepts~\cite{NuDoubt_2025}.

\subsection{Hybrid Slow Scintillators}
In recent years, novel ideas in the field of liquid scintillator technology emerged –~most prominently hybrid and opaque scintillators~– which both allow for the first time a discrimination between electron, gamma-rays, and positrons at MeV-scale~\cite{Schoppmann_2023}.
Hybrid scintillators exploit the particle-dependent ratio of prompt Cherenkov and delayed scintillation light (\v{C}/S ratio) to discriminate between particles or event types~\cite{Steiger_2024,Theia_2019}.
Particles close to the Cherenkov threshold produce less Cherenkov light.
While electrons tend to have a larger ratio, gamma-rays produce lesser amounts of Cherenkov light, as they transfer only small amounts of energy into several recoil-electrons below or close to the threshold.
For positrons, the relative amount of Cherenkov light is even lower with respect to a gamma-ray, because the total amount of visible scintillation light $E_\text{vis}$ in a detector includes $2 \cdot 511\,\text{keV}$ energy from two annihilation gammas, which produce no Cherenkov light, as explained above.
Therefore, the kinetic energy of the positron is much lower than the overall energy of the event~\cite{NuDoubt_2025}.
This effect is even stronger for double positron events.
The hybrid discrimination strategy allows to separate signal and background events for the first time into distinct populations and allows \nudoubt~to excel at background suppression.

\subsection{Opaque Scintillators}
\begin{figure}[tb]
\centering
    \begin{subfigure}[b]{0.24\textwidth}
    \centering
        \caption{}
        \vspace{-0.3cm}
        \includegraphics[width=\textwidth]{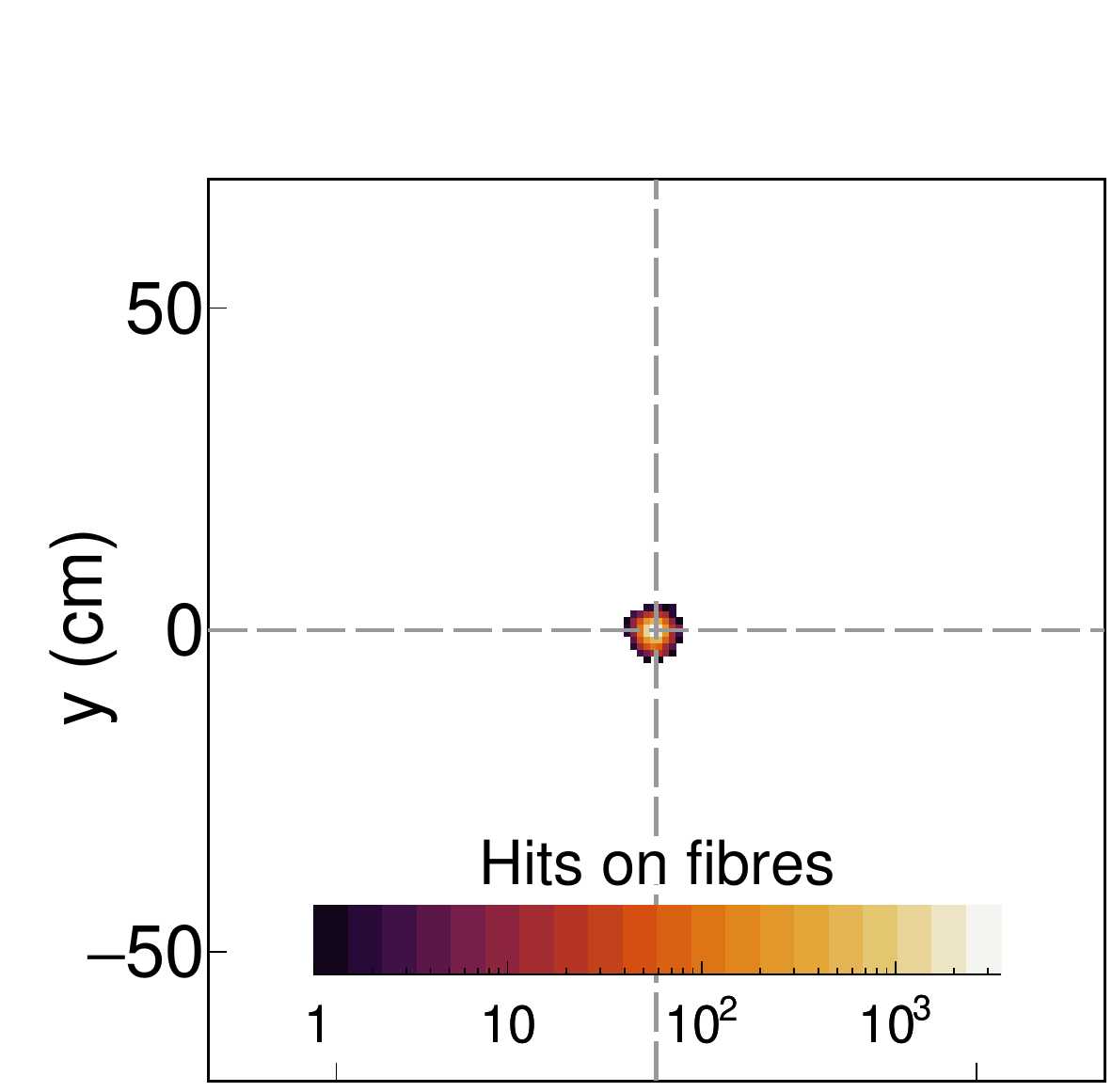}
        \label{fig:subfig_a}
    \end{subfigure}
    \vspace{-0.3cm}
    \hspace{-0.2cm}
    \begin{subfigure}[b]{0.24\textwidth}
    \centering
        \caption{}
        \vspace{-0.3cm}
        \includegraphics[width=\textwidth]{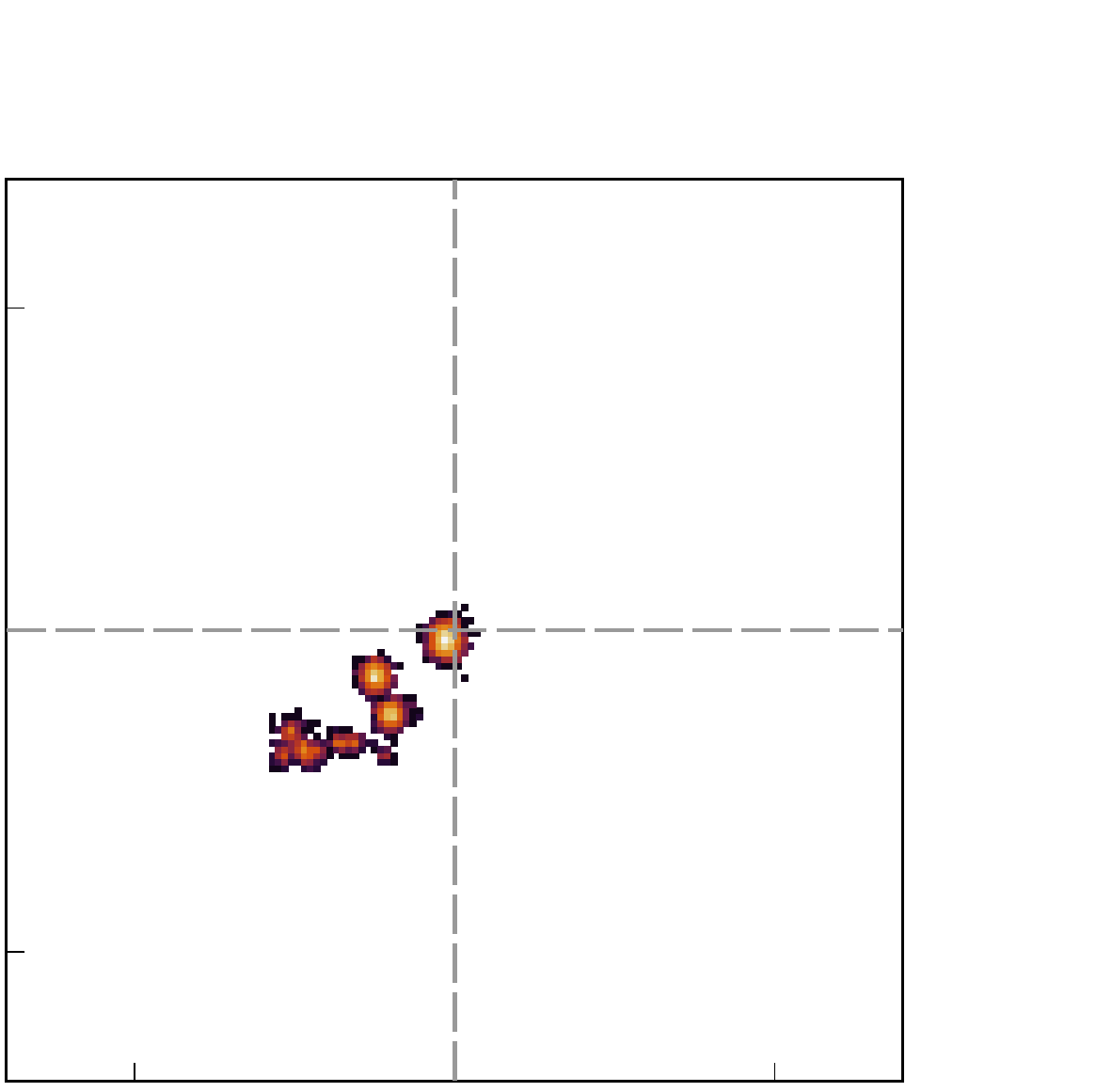}
        \label{fig:subfig_b}
    \end{subfigure}
    \begin{subfigure}[b]{0.24\textwidth}
    \centering
        \caption{}
        \vspace{-0.3cm}
        \includegraphics[width=\textwidth]{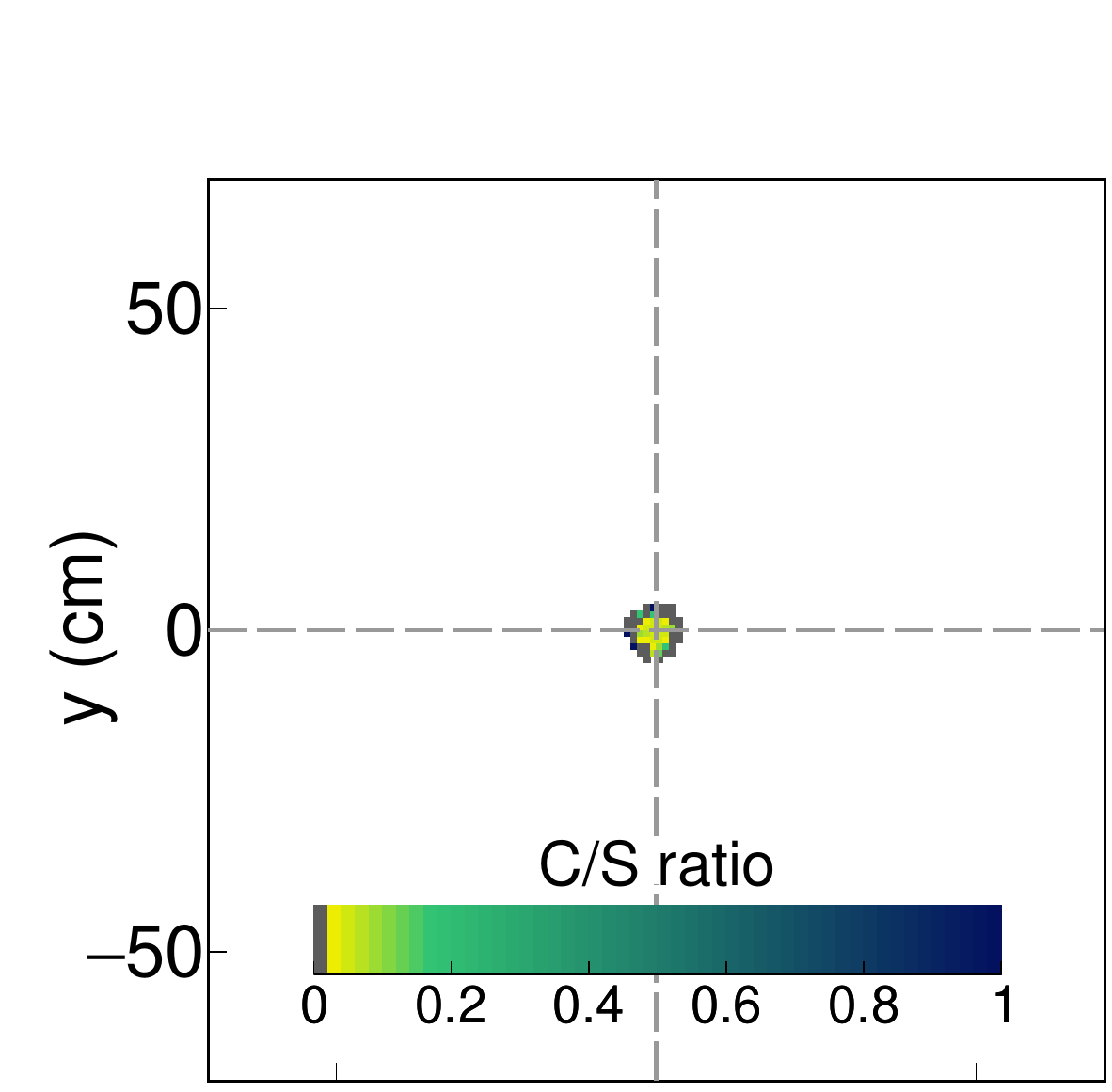}
        \label{fig:CSsubfig_a}
    \end{subfigure}
    \vspace{-0.3cm}
    \hspace{-0.2cm}
    \begin{subfigure}[b]{0.24\textwidth}
    \centering
        \caption{}
        \vspace{-0.3cm}
        \includegraphics[width=\textwidth]{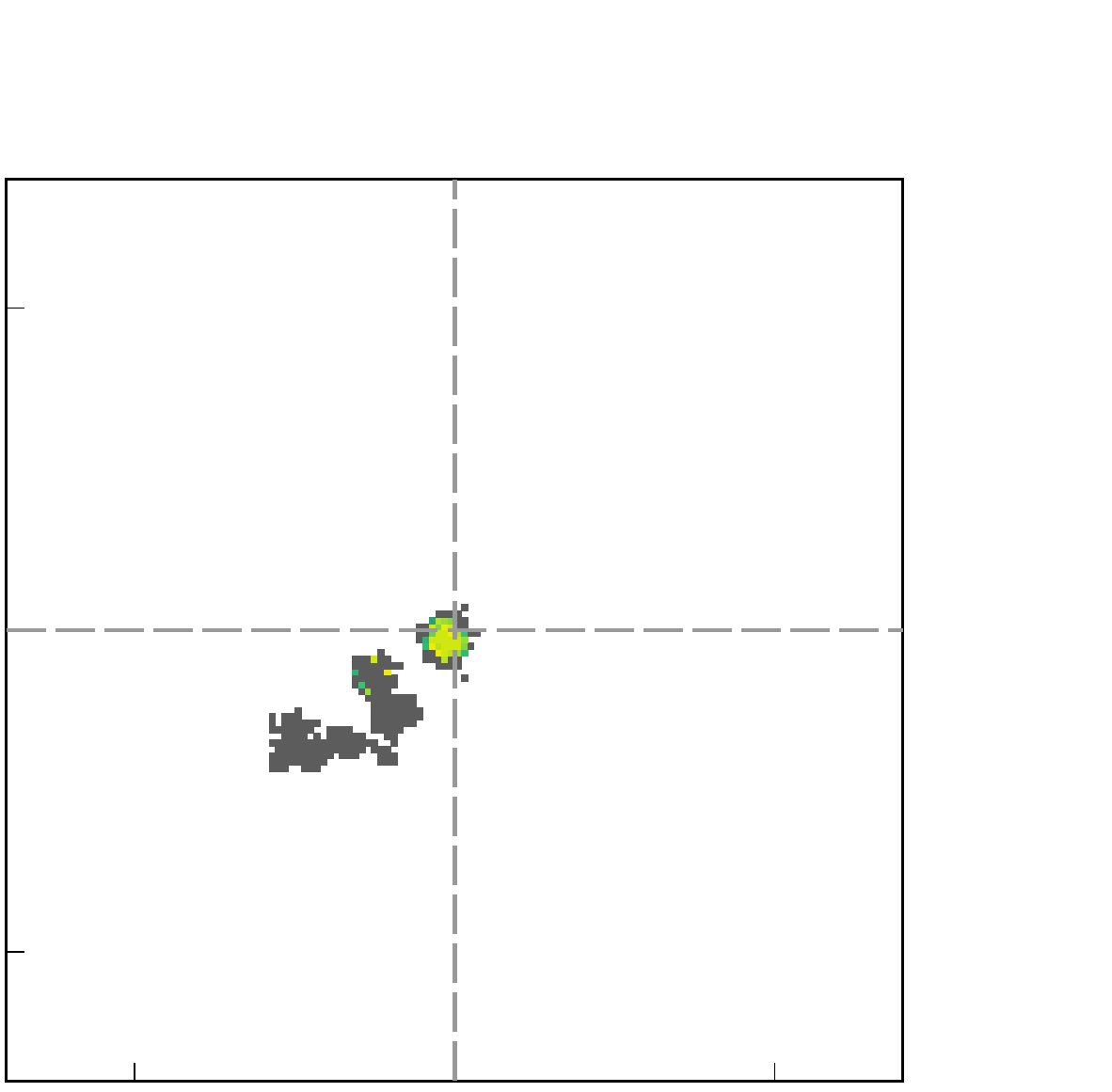}
        \label{fig:CSsubfig_b}
    \end{subfigure}
    \begin{subfigure}[b]{0.24\textwidth}
    \centering
        \includegraphics[width=\textwidth]{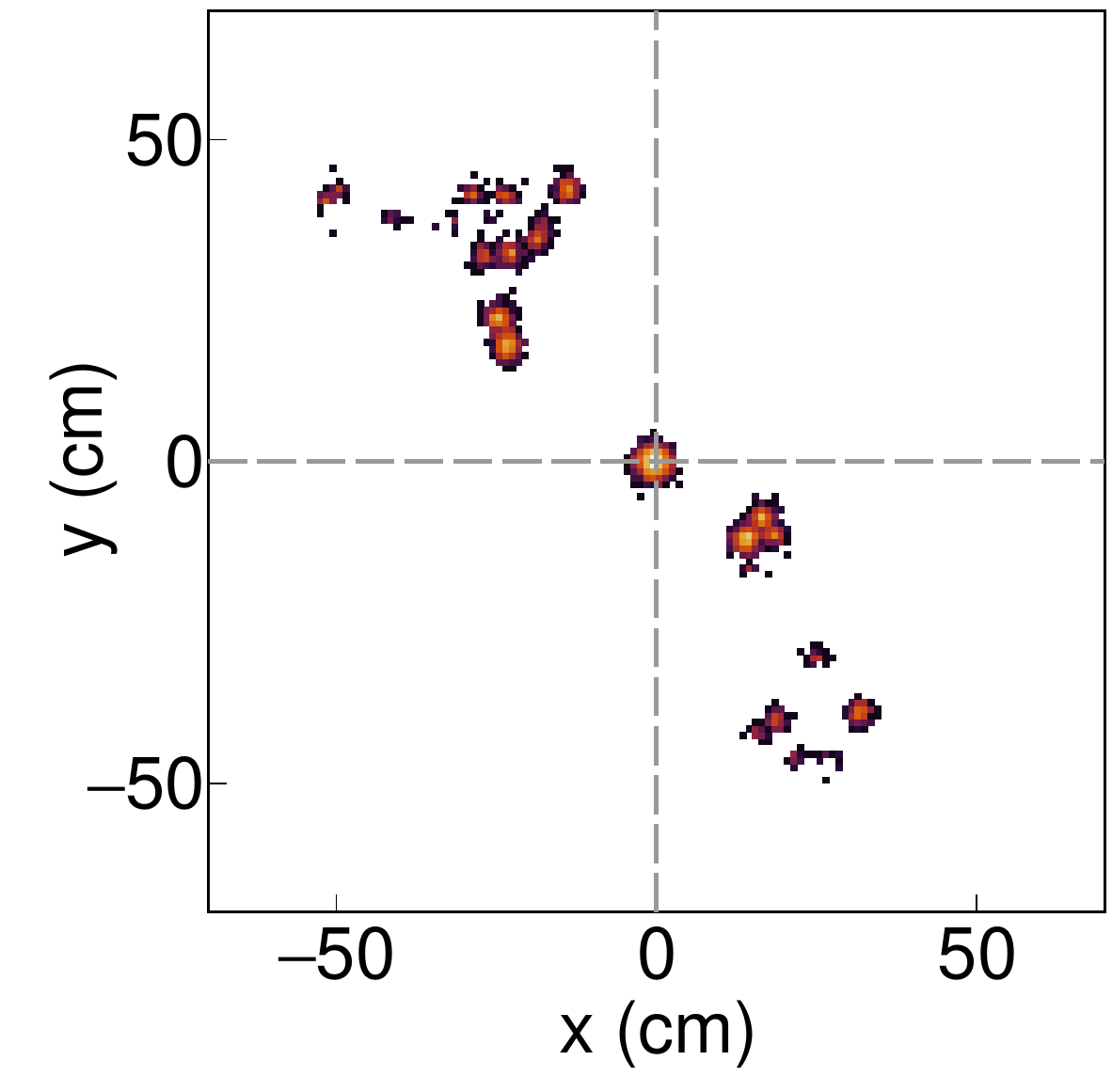}
        \caption{}
        \label{fig:subfig_c}
    \end{subfigure}
    \hspace{-0.2cm}
    \begin{subfigure}[b]{0.24\textwidth}
    \centering
        \includegraphics[width=\textwidth]{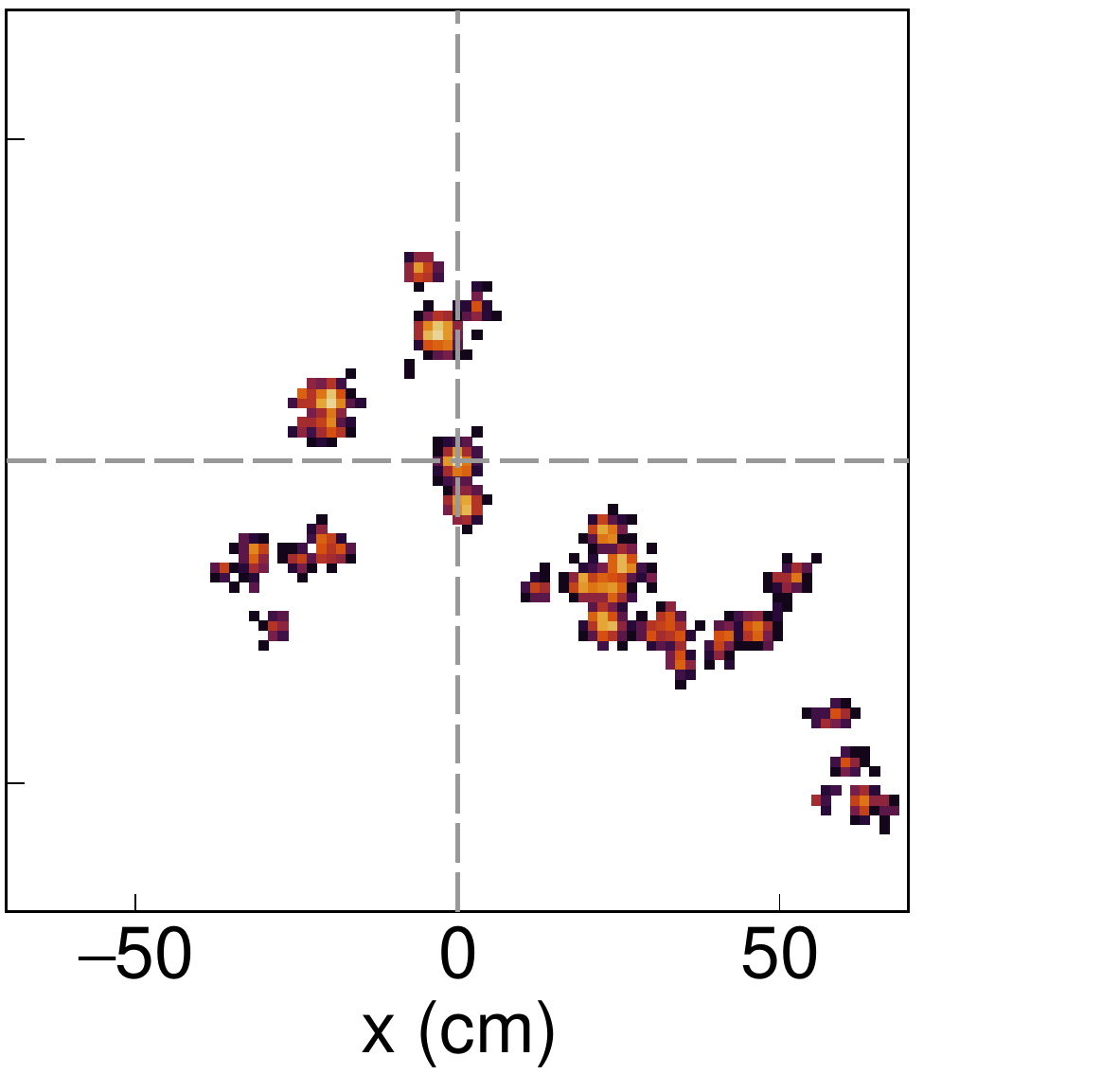}
        \caption{}
        \label{fig:subfig_d}
    \end{subfigure}
    \begin{subfigure}[b]{0.24\textwidth}
    \centering
        \includegraphics[width=\textwidth]{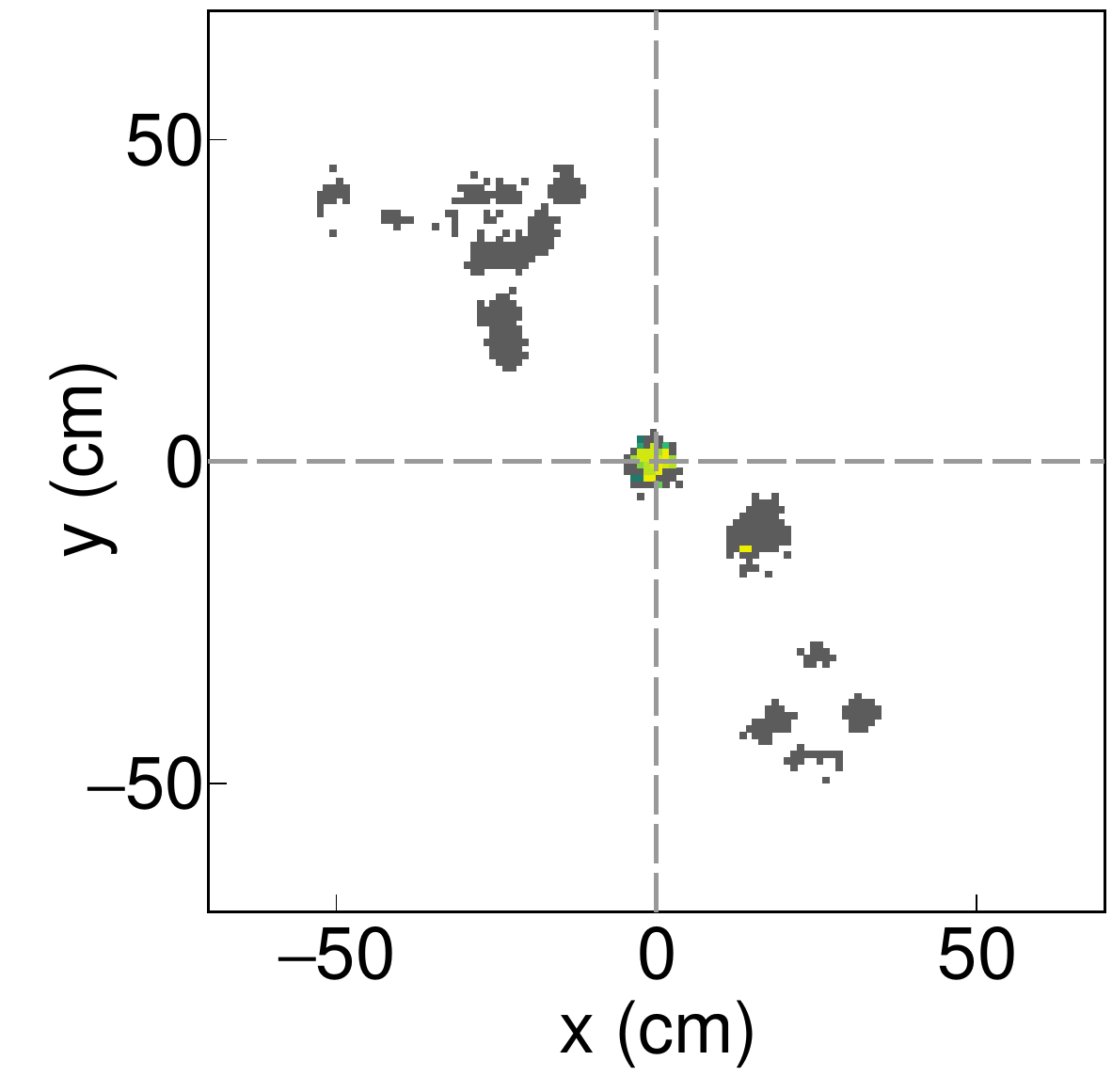}
        \caption{}
        \label{fig:CSsubfig_c}
    \end{subfigure}
    \hspace{-0.2cm}
    \begin{subfigure}[b]{0.24\textwidth}
    \centering
        \includegraphics[width=\textwidth]{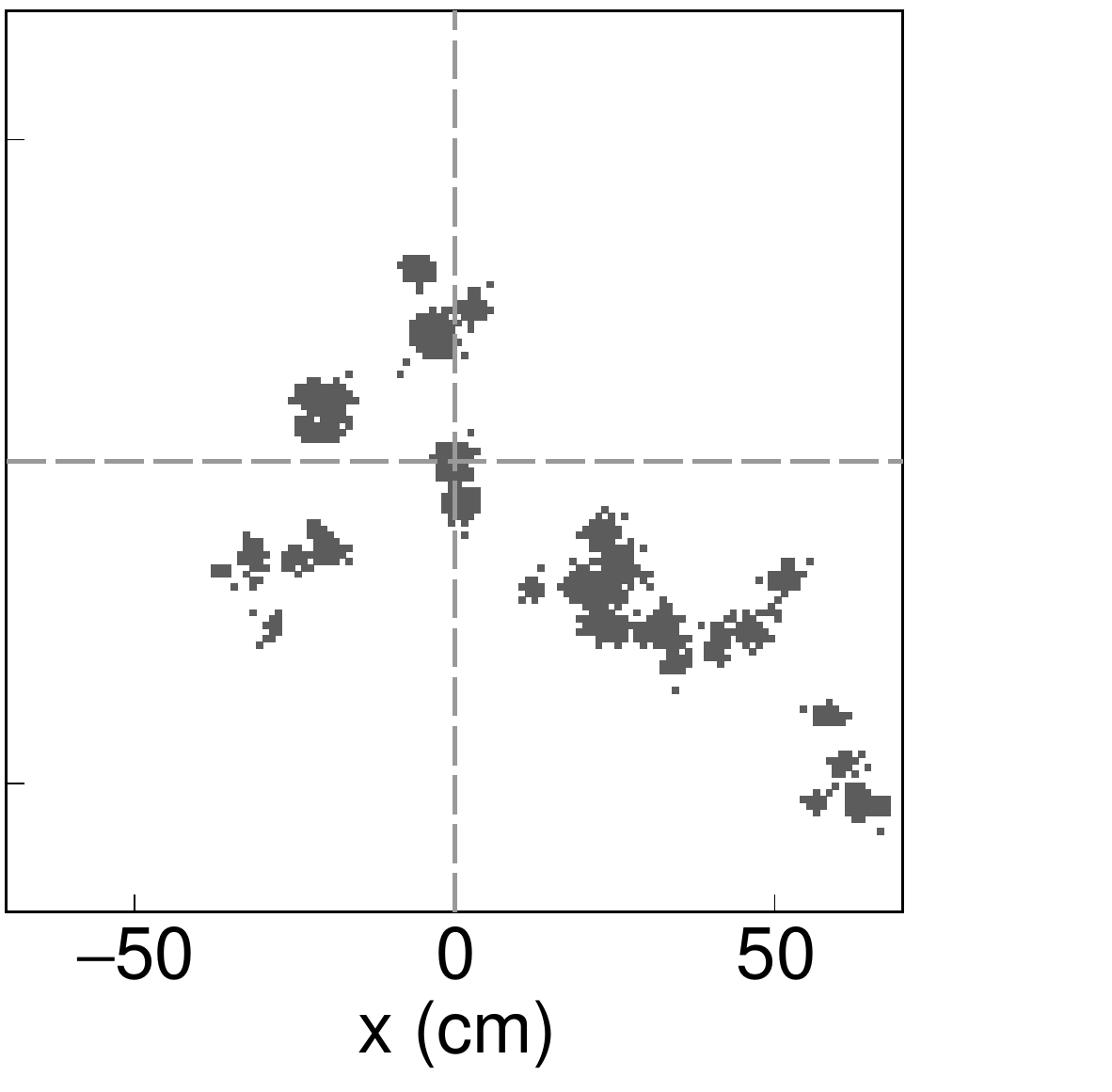}
        \caption{}
        \label{fig:CSsubfig_d}
    \end{subfigure}
    \caption{Illustration of the proposed strategy of particle discrimination in opaque scintillator based on the morphology of hits of optical photons per fibre. Geant4~\cite{Geant4_2003} simulations of (\ref{fig:subfig_a}) an electron, (\ref{fig:subfig_b}) a gamma-ray, (\ref{fig:subfig_c}) a single positron, and (\ref{fig:subfig_d}) two positrons. All four events deposit a total energy of 2.8\,MeV in the detector. Additionally, the ratio of hits of Cherenkov and scintillation photons for each fibre is displayed (\ref{fig:CSsubfig_a}, \ref{fig:CSsubfig_b}, \ref{fig:CSsubfig_c}, and \ref{fig:CSsubfig_d}) for the same four events.
    Hit fibres with $\text{\v{C}/S} = 0$ are represented in grey on the colour scale.
    Fibres without hits have white colour. Figures reproduced from~\cite{NuDoubt_2025} under \href{http://creativecommons.org/licenses/by/4.0/}{CC BY 4.0 license}.}
    \label{fig:blobs}
\end{figure}

In \nudoubt, we plan to introduce opacity to the hybrid scintillator, which offers high spatial resolution and introduces additional possibilities for particle identification via morphological patterns.
Opaque scintillators confine scintillation light through Mie-Lorentz scattering of optical photons~\cite{Buck_2019,Schoppmann_2026,Schoppmann_2025_1,Schoppmann_2025_2,LiquidO_2021}.
The local light depositions are then picked up by a dense grid of wavelength-shifting fibres.
The fibres are running through the entire detector volume and guide the light to both their ends, where it is read out with silicon photomultipliers (SiPMs).
The opaque detector technology allows groundbreaking particle identification power, because it retains the morphological information, where ionising particles have deposited their energy, because the scintillation light stays close to the interaction points of the particles.
As can be seen in \autoref{fig:blobs}, electrons deposit their energy in a short ionisation trail and create a single blob (\textbf{b}ulky \textbf{l}ight \textbf{ob}ject) of scintillation light of a few centimetres size, gamma-rays leave several blobs as they do multiple Compton-scatters followed by a final photoelectric effect.
Positrons combine the patterns of electrons and gammas, because they first lose their kinetic energy in an ionisation trail like an electron and then produce two annihilation gamma-rays, which leave their characteristic patterns.
Double-positron events are an overlay of two positron events.
As shown in \autoref{fig:blobs}, the combination of opaque and hybrid scintillator allows to clearly identify the blobs with Cherenkov light and thereby allows to gain additional discrimination power beyond the opaque-only and hybrid-only techniques, based on the position of the Cherenkov blob relative to the other blobs.

\subsection{Optimised Light Guides}
The largest loss of light happens at the interfaces between scintillator and wavelength-shifting fibres, as the geometrical capture efficiency via total internal reflection of the isotropically emitted and wavelength-shifted light is under 10\%.
\nudoubt~will overcome this loss by using optimised wavelength-shifting light guides (OWL-fibres~\cite{Kessler_2024_1,Kessler_2024_2}).
To maximise light capture of the OWL-fibres, the wavelength-shifters (WLS) are located on the surface of the fibre since photons that are absorbed and emitted there have a higher chance of being captured by total internal reflection compared to those closer to the centre of the fibre~\cite{Bastian-Querner_2022}.
The OWL-fibres feature an almost five times increase capture efficiency.
Thus, a light yield of more than 800\,PE/MeV appears possible in \nudoubt, an improvement by a factor of two compared to the current state of the art~\cite{LiquidO_2024}.
This corresponds to an energy resolution of about 2\% in the region of interest (ROI) for positron-emitting neutrinoless double weak decays~\cite{NuDoubt_2025}.
We have manufactured OWL-fibres with millimetre-range diameters which we dip coated with bis-methylstyrylbenzene (bis-MSB) as wavelength shifter.
These fibres have an attenuation length of 2\,m fully sufficient for usage in our compact detector concept.

\subsection{Loading of Isotopes}
\nudoubt~will be loaded with 50\% enriched krypton-78 gas under 5 bar over-pressure.
This new approach of pressure loading allows to dissolve 5 times more gas than in previous experiments, allowing \nudoubt~to reach unprecedented signal rates for a scintillation detector~\cite{Eisenhuth_2024}.
The energy release of krypton-78 is well above the gamma background of thallium, allowing \nudoubt~to reach minimal background rates~\cite{NuDoubt_2025}.
We expect that pressure-loading has negligible effect on the optical properties of our scintillator.
For verification, we constructed an 100\,ml overpressure test cell, designed to determine the transparency, light yield, and loading factor of our liquid scintillator as function of pressure~\cite{Eisenhuth_2024}.

After conclusion of the krypton phase, the krypton-78 gas can be exchanged with xenon-124, another nobel gas with similarly high decay energy.
After the conclusion of the two gas loading phases, cadmium-106 can be homogeneously dispersed as cadmium tungstate powder throughout the viscous hybrid-opaque scintillator.
Cadmium-106 also features a high decay energy above the thallium background.
The tungstate crystals are self-scintillating, therefore the loading does not degrade scintillator light yield~\cite{Schoppmann_2026,NuDoubt_2025}.
Finally, it is possible to replace the cadmium-tungstate by barium-sulfate powder.
Just like cadmium-tungstate powder, barium-sulfate powder aids the opacity of the scintillator as it is highly reflective and provides additional scattering targets~\cite{Schoppmann_2026,Schoppmann_2025_1,Schoppmann_2025_2}.
However, since barium-sulfate is not scintillating, the barium-loaded scintillator will exhibit a lower light yield.
Nevertheless, it will be possible to probe positive double weak decays of barium-130.
\nudoubt~will thus be able to probe four of the six isotopes that can undergo all three positive double weak decay modes~\cite{Palusova_2025}.

\section{Detector Design}
\nudoubt~is going to use a compact cylindrical tonne-scale scintillation detector of about 150\,cm height and diameter~\cite{NuDoubt_2025}.
The centre of the detector will consist of a vessel filled with a hybrid-opaque liquid scintillator, pressure-loaded with 50\% enriched krypton-78 gas.
The central vessel will operate at 5\,bar overpressure and contains approximately 10\,kg of active scintillating material.
To detect Compton-scatters from annihilation gamma-rays, the central vessel is surrounded by an unloaded hybrid-opaque scintillator volume. Both hybrid-opaque scintillator volumes are going to be instrumented with a grid of unilateral optimized wavelength-shifting (OWL) fibres, as detailed above.
These fibres run along the symmetry axis of the cylindrical detector and are read out at both ends by silicon photomultipliers (SiPMs), allowing for three-dimensional reconstruction of the event topology using fibre position and timing delay.
A third vessel containing a transparent liquid scintillator and instrumented with photomultiplier tubes (PMTs) is going to surround the central two vessels.
This layer serves as shielding and as active veto system to identify and reject external backgrounds such as cosmic muons or gamma-radiation from surrounding materials.
This outer volume could also be realised as a water-Cherenkov detector.

\section{Expected Sensitivities}
Sensitivity studies were carried out based on data from several table-top prototypes.
These prototypes verify key aspects of the technology such as scintillator light yield, scattering length, separation of Cherenkov and scintillation light, coupling between scintillator, fibres and SiPMs, pressure-loading, as well as fibre production and performance.
Assuming conservative cleanliness levels and background rejection power and assuming further a cosmic muon flux at the level of the Gran Sasso or Boulby underground laboratory, we expect to measure for the first time positron-emitting double weak decays within less than two calendar years of operation~\cite{NuDoubt_2025}.
Within the same exposure, we further expect a background-free search in the ROI for neutrinoless decays and thereby improve limits by almost three orders of magnitude.

\section{Project Timeline}
We are currently testing our detector concept in an upscaled detector prototype with 250 fibres that has a volume of approximately 30 litres~\cite{Girard-Carillo_2025}.
We further plan to conclude construction of the full detector prototype within the next three years.
After an commissioning and engineering run on surface level, in which we verify the pressure-loading and the hybrid-opaque particle discrimination by means of deployment of radioactive sources, we intend to bring the detector prototype to an underground facility.
We are then going to perform the physics measurement using krypton-78 in the following two years.
This will be followed by physics measurements of the additional double beta plus isotopes xenon-124, cadmium-106, barium-130.
Further isotopes are currently explored for loading.

\section{Conclusion}
The \nudoubt~experiment is going to exploit four novel detector concepts in the search for positron-emitting double weak decays, combining hybrid-slow and opaque scintillator technologies with advanced light collection using OWL fibers and pressure-loading of nobel gas.
A prototype is under active development, with promising sensitivities towards Standard Model and neutrinoless modes~\cite{NuDoubt_2025}.
We plan to perform a first physics measurement within the next five years.

\section*{Acknowledgements}
This work has been supported by the Cluster of Excellence ``Precision Physics, Fundamental Interactions, and Structure of Matter'' (PRISMA$^{+}$ EXC 2118/1) funded by the German Research Foundation (DFG) within the German Excellence Strategy (Project ID 390831469).

\end{document}